\documentclass[pdflatex,sn-mathphys-num]{sn-jnl}

\usepackage{graphicx}%
\usepackage{multirow}%
\usepackage{amsmath,amssymb,amsfonts}%
\usepackage{amsthm}%
\usepackage{mathrsfs}%
\usepackage[title]{appendix}%
\usepackage{xcolor}%
\usepackage{textcomp}%
\usepackage{manyfoot}%
\usepackage{booktabs}%
\usepackage{algorithm}%
\usepackage{algorithmicx}%
\usepackage{algpseudocode}%
\usepackage{listings}%
\usepackage[colorinlistoftodos,prependcaption,textsize=tiny]{todonotes}
\theoremstyle{thmstyleone}%
\theoremstyle{thmstyletwo}%

\theoremstyle{thmstylethree}%
\begin{document}

\title[End-to-End Reverse Screening Identifies Protein Targets of Small Molecules Using HelixFold3]{End-to-End Reverse Screening Identifies Protein Targets of Small Molecules Using HelixFold3}


\author[1,2]{\fnm{Shengjie} \sur{Xu}}\email{u202317280@hust.edu.cn}

\author[1]{\fnm{Xianbin} \sur{Ye}}\email{yexianbin@baidu.com}

\author[3]{\fnm{Mengran} \sur{Zhu}}\email{u202313729@hust.edu.cn}

\author[1]{\fnm{Xiaonan} \sur{Zhang}}\email{zhangxiaonan@baidu.com}

\author*[1]{\fnm{Shanzhuo} \sur{Zhang}}\email{zhangshanzhuo@baidu.com}

\author*[1]{\fnm{Xiaomin} \sur{Fang}}\email{fangxiaomin01@baidu.com}

\affil[1]{\orgdiv{PaddleHelix Team}, \orgname{Baidu Inc.}, \city{Shenzhen}, \country{China}}

\affil[2]{\orgdiv{School of Software Engineering}, \orgname{Huazhong University of Science and Technology}, \city{Wuhan}, \country{China}}

\affil[3]{\orgdiv{School of Pharmacy, Tongji Medical College}, \orgname{Huazhong University of Science and Technology}, \city{Wuhan}, \country{China}}


\abstract{Identifying protein targets for small molecules, or reverse screening, is essential for understanding drug action, guiding compound repurposing, predicting off-target effects, and elucidating the molecular mechanisms of bioactive compounds. Despite its critical role, reverse screening remains challenging because accurately capturing interactions between a small molecule and structurally diverse proteins is inherently complex, and conventional step-wise workflows often propagate errors across decoupled steps such as target structure modeling, pocket identification, docking, and scoring.

Here, we present an end-to-end reverse screening strategy leveraging HelixFold3, a high-accuracy biomolecular structure prediction model akin to AlphaFold3, which simultaneously models the folding of proteins from a protein library and the docking of small-molecule ligands within a unified framework. We validate this approach on a diverse and representative set of approximately one hundred small molecules. Compared with conventional reverse docking, our method improves screening accuracy and demonstrates enhanced structural fidelity, binding-site precision, and target prioritization. By systematically linking small molecules to their protein targets, this framework establishes a scalable and straightforward platform for dissecting molecular mechanisms, exploring off-target interactions, and supporting rational drug discovery.}




\makeatletter
\gdef\authemail{}%
\makeatother

\maketitle

\newpage

\section{Introduction}

\begin{figure}[t]
    \centering
    \includegraphics[width=1.0\linewidth]{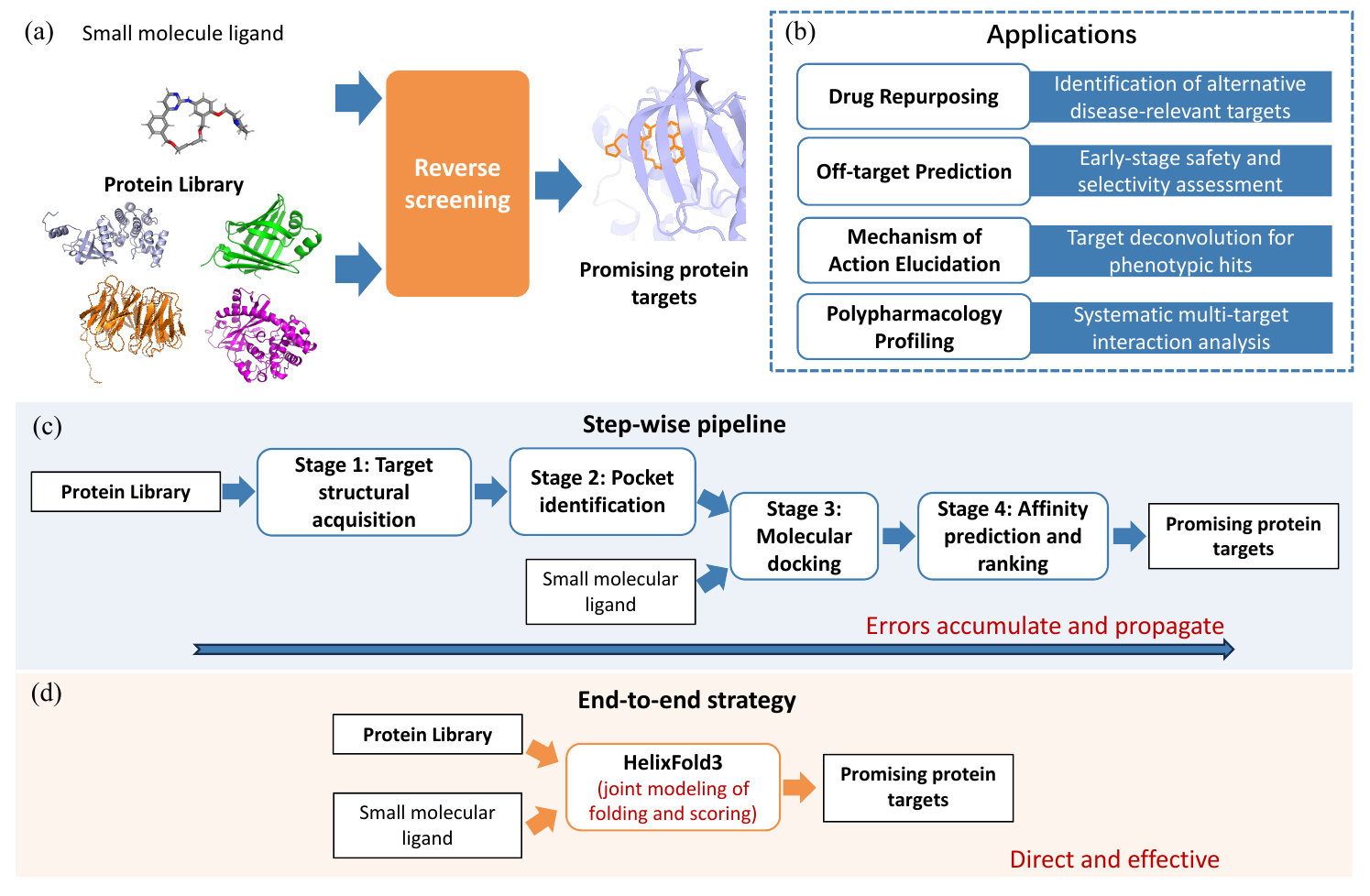}
    \caption{\textbf{Computational paradigms and applications of reverse screening.} \textbf{(a,b)} Concept and therapeutic applications of reverse screening. \textbf{(c)} Conventional step-wise pipeline, highlighting the accumulation and propagation of errors across decoupled steps. \textbf{(d)} HelixFold3 end-to-end strategy, which unifies protein folding and ligand scoring into a single, direct predictive process.}
    \label{fig:framework}
\end{figure}

Reverse screening \cite{huang2018reverse} constitutes a pivotal computational paradigm for deciphering the multi-target profiles of small molecules \cite{keiser2009predicting,huang2018review}. Distinct from conventional forward screening, which explores chemical space for a predefined protein target, reverse screening (Fig.~\ref{fig:framework}a) systematically maps the ligand-protein interactome by prioritizing candidate targets based on binding energetics and structural complementarity. As illustrated in Fig.~\ref{fig:framework}b, reverse screening underpins several key applications in drug discovery. It enables drug repurposing through the identification of alternative disease-relevant targets, supports early-stage off-target prediction by assessing compound selectivity and safety risks, facilitates mechanism of action (MoA) elucidation for phenotypic screening hits, and provides a systematic framework for polypharmacology profiling by characterizing multi-target interaction patterns.

Despite its transformative potential, the robust implementation of reverse screening remains a formidable challenge. The primary difficulty stems from the requirement to precisely model intricate physicochemical determinants across an expansive and structurally heterogeneous protein space. Unlike conventional forward screening, where the search is confined to a predefined binding site, reverse screening must navigate highly variable three-dimensional topologies, diverse surface chemistries, and complex conformational dynamics at a proteome-wide scale. Consequently, this field remains relatively underexplored.

To identify potential protein targets for small molecules, computational strategies generally fall into two categories: ligand-centric and structure-based approaches. While ligand-centric methods leveraging chemical similarity, pharmacophore matching, or machine learning offer high-throughput capabilities \cite{schuffenhauer2003similarity,wang2024comet,daina2024testing}, they typically lack the mechanistic resolution required for de novo discovery as they bypass explicit modeling of molecular interactions. Recent large-scale evaluations have demonstrated that machine learning-enhanced reverse screening can achieve promising target identification rates \cite{daina2024testing}, and multi-module approaches combining ligand similarity, docking, and AI affinity prediction have shown potential for balancing speed and accuracy \cite{wang2024comet}. Consequently, structure-based reverse docking has emerged as the primary paradigm for mechanistic target fishing \cite{kharkar2014reverse,lee2016using,luo2024benchmarking}. However, traditional implementations of this paradigm rely on a step-wise pipeline involving protein structural acquisition, pocket identification, molecular docking, and affinity scoring prediction, which is inherently prone to error propagation (Fig.~\ref{fig:framework}c).
The limitations of this approach appear at every stage of the process. In the initial structural phase, static conformations often fail to represent the flexibility required to accommodate different ligands. During pocket identification, algorithms frequently pick out surface cavities that have no biological relevance to actual binding events. Within the docking stage, errors in side-chain positioning and the failure to account for induced-fit changes can significantly reduce the accuracy of predicted binding poses. Finally, in the scoring phase, the reliance on these imprecise docking structures results in a weakened screening capacity, as scoring functions struggle to distinguish true targets from decoys when the underlying binding geometry is flawed. As a result, errors accumulate and propagate across the reverse docking pipeline. Ultimately, neither ligand-centric models nor step-wise structure-based pipelines can fully capture the concerted nature of molecular recognition, highlighting an urgent need for end-to-end, structure-based frameworks that can unify protein folding and ligand binding within a single, integrated predictive process.

To overcome these challenges, we developed an end-to-end, structure-based reverse screening strategy that treats protein folding, ligand docking, and affinity estimation as a jointly optimized process (Fig.~\ref{fig:framework}d). By unifying these previously decoupled stages, this framework bypasses the need for intermediate steps, thereby mitigating the systematic error accumulation inherent in traditional workflows. This strategy is implemented using HelixFold3  \cite{liu2024technicalreporthelixfold3biomolecular}, a high-accuracy biomolecular structure prediction model capable of simultaneously modeling the full complex of proteins from a protein library and small-molecule ligands. By taking only protein sequences and ligand SMILES as inputs, HelixFold3 directly predicts protein-ligand complex geometries while generating intrinsic confidence metrics, including interface predicted Template Modeling score \cite{evans2021alphafoldmultimer} (ipTM) and predicted Local Distance Difference Test \cite{jumper2021highly,mariani2013lddt} (pLDDT) of the ligand, to rank proteins from the protein library.

To investigate the capacity of our end-to-end strategy, we first validated its general reverse screening performance across a diverse library of approximately one hundred representative protein-ligand pairs with available co-crystal structures. For each ligand, the model was tasked with identifying its native binding partner from a candidate pool of approximately 1,000 human proteins, which were systematically sampled from the human proteome via sequence clustering to ensure diversity. Compared with conventional step-wise reverse docking pipelines, our approach demonstrates significantly improved structural fidelity and a remarkable enhancement in screening performance, with the success rate increased by several folds. This performance leap is underpinned by the superior accuracy of the end-to-end approach across all dimensions, including target structure prediction, pocket identification, ligand docking, and affinity scoring. Beyond general benchmarking, we further evaluated the platform's utility in specialized scenarios, including drug repurposing and off-target prediction. To this end, we curated a high-quality evaluation dataset consisting of two clinically relevant drugs: one representing a classic drug repurposing scenario and one exemplifying off-target effect prediction, each with their experimentally verified targets. By challenging HelixFold3 to recover these complex associations, we provide evidence of its practical efficacy. These results illustrate the potential of our strategy as a robust solution for investigating molecular mechanisms and guiding rational chemical biology studies.

\section{Results}

\subsection{Benchmarking the End-to-End Reverse Screening Paradigm}

\begin{figure}[t]
    \centering
    \includegraphics[width=1.0\linewidth]{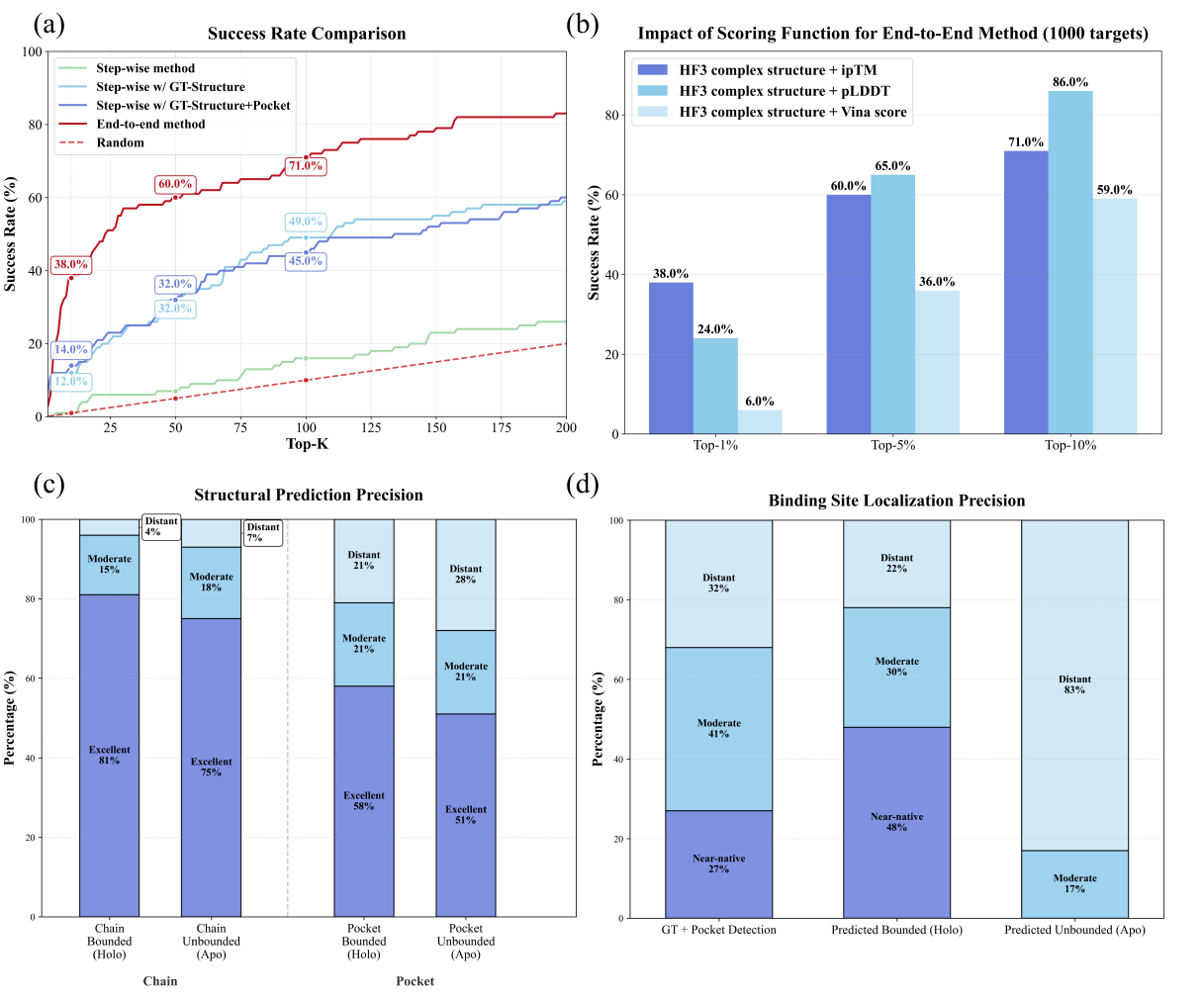}
    \caption{\textbf{Comprehensive performance evaluation of reverse screening methods.} \textbf{(a)} Success rate comparison between step-wise method and our end-to-end method. \textbf{(b)} Impact of scoring function for the end-to-end method. \textbf{(c)} Structural prediction precision of the whole chain or pocket of proteins. The excellent, moderate, and distant groups are classified by the structure difference between predicted and crystallized protein structures (holo or apo). \textbf{(d)} Binding site localization precision comparing GT + pocket detection, predicted bounded (holo), and predicted unbounded (apo). }
    \label{fig:fig2}
\end{figure}

To evaluate the efficacy of our end-to-end framework, we established a rigorous reverse screening benchmark described in Section~\ref{sec:datasets} (Evaluation Datasets). We compared our HelixFold3-based end-to-end approach against three step-wise baseline pipelines (Fig.~\ref{fig:fig2}a). The standard step-wise pipeline (light green line in Fig.~\ref{fig:fig2}a) represents traditional modular workflows: structures for the 100 native proteins were predicted using HelixFold3, while the 900 decoys utilized pre-calculated structures from the AlphaFold protein structure database \cite{alphafolddb2022} (AFDB). Pocket identification followed a hierarchical protocol: candidate sites were primarily identified using P2Rank \cite{krivak2018p2rank}, with fpocket \cite{le2009fpocket} serving as a fallback when no high-confidence pockets were detected. Molecular docking and affinity scoring were subsequently performed using AutoDock Vina \cite{trott2010autodock,vina2021}. To further dissect the performance bottlenecks of traditional methods, we constructed two additional step-wise pipeline variants: (1) Step-wise w/ GT-Structure (light blue line in Fig.~\ref{fig:fig2}a): Utilized ground-truth (GT) native protein structures directly from the PDB, while maintaining the P2Rank/fpocket-based pocket detection. (2) Step-wise w/ GT-Structure+Pocket (medium blue line in Fig.~\ref{fig:fig2}a): Utilized both the ground-truth PDB structures and the exact pocket coordinates derived from the co-crystal ligands. In contrast to these modular approaches, our HelixFold3-based end-to-end framework (red line in Fig.~\ref{fig:fig2}a) operates as a jointly optimized process. It requires only protein sequences and ligand SMILES as direct inputs, simultaneously modeling the full complex geometry and generating intrinsic ipTM and pLDDT confidence metrics for ranking.

As shown in Fig.~\ref{fig:fig2}a, the HelixFold3-based end-to-end approach (red line) significantly outperforms all conventional step-wise pipelines across the entire ranking range. Notably, the end-to-end approach achieved a Top-1\% success rate of 38.0\% and a Top-10\% success rate of 71.0\%. These figures represent a several-fold improvement over the various step-wise pipelines, even when those baselines were granted access to crystal protein structures and pocket coordinates. This performance gap underscores the robust capability of the end-to-end strategy in reverse screening. A detailed ablation analysis through these baselines reveals the critical factors governing screening performance. First, the comparison between Step-wise w/ GT-Structure (light blue line) and the standard step-wise pipeline (light green line) highlights that the accuracy of the ligand-bound (holo) protein structure plays a pivotal role in successful ranking. Second, the observation that Step-wise w/ GT-Structure+Pocket (medium blue line) performs similarly to Step-wise w/ GT-Structure suggests that current pocket identification tools are relatively effective and are not the primary bottleneck in this benchmark. Finally, the substantial superiority of the end-to-end approach over the Step-wise w/ GT-Structure+Pocket variant indicates that ligand docking and affinity scoring accuracy are also vital for correct ranking.

We investigated the impact of different scoring functions on the screening success rate within the end-to-end framework (Fig.~\ref{fig:fig2}b). We evaluated three scoring metrics: ipTM, ligand pLDDT, and the Vina score applied to HelixFold3-predicted complex structures. While the Vina score yielded the lowest performance (6.0\% success rate at Top-1\%, 36.0\% at Top-5\%, and 59.0\% at Top-10\%), the intrinsic confidence metrics of HelixFold3 proved significantly more effective. Notably, ipTM exhibited superior early enrichment at the Top-1\% level (38.0\%), while the ligand-specific pLDDT reached the highest overall success rate at broader thresholds, achieving 24.0\% at Top-1\%, 65.0\% at Top-5\%, and 86.0\% at Top-10\% (Fig.~\ref{fig:fig2}b). These results demonstrate that the model's internal confidence scores are better calibrated for the co-folded interfaces than traditional physical-based scoring functions, further consolidating the advantage of the end-to-end paradigm in identifying true protein-ligand pairs. 

\subsection{Determinants of Superior Screening Performance}
To understand why the end-to-end strategy excels, we first evaluated the structural consistency between the HelixFold3-predicted holo-protein structures and their corresponding unbound (apo) structures, using the ground-truth holo structures from the PDB \cite{berman2000protein} as a reference. We computed RMSD values at two levels: chain-level RMSD measures the structural deviation of the entire target protein chain relative to the ground-truth structure, while pocket-level RMSD focuses specifically on the binding pocket region, which is functionally more relevant to ligand binding. We categorized RMSD values into three layers: Excellent (RMSD $<$ 0.5 \AA), Moderate (0.5 $\le$ RMSD $<$ 1.0 \AA), and Distant (RMSD $\ge$ 1.0 \AA). For chain-level RMSD, the predicted holo structures (Chain Bounded) achieved 81.0\% in the Excellent category, compared to 75.0\% for predicted apo structures (Chain Unbounded) (Fig.~\ref{fig:fig2}c). For pocket-level RMSD, the predicted holo structures (Pocket Bounded) achieved 58.0\% in the Excellent category, compared to 51.0\% for predicted apo structures (Pocket Unbounded) (Fig.~\ref{fig:fig2}c). This increased structural fidelity within the binding pocket—specifically the model's ability to capture induced-fit rearrangements—underpins the superior ranking performance by providing a more accurate predicted ligand-bound structure.
Then, we analyzed the impact of binding site localization quality (Fig.~\ref{fig:fig2}d). For binding site localization, we categorized the distance between predicted binding site centers and ground-truth ligand centers—defined as the centroid of the ligand coordinates extracted from the experimental PDB \cite{berman2000protein} co-crystal structure—into three layers: Near-native ($\le2$ \AA), Moderate ($2$--$8$ \AA), and Distant ($\ge8$ \AA). The method for determining the predicted center varied by approach, with the end-to-end strategy utilizing the centroid of the ligand coordinates directly predicted by HelixFold3, while the step-wise pipelines utilized the centroid of the highest-scoring pocket identified by P2Rank or fpocket. Our analysis reveals that HelixFold3-predicted binding sites (Predicted Bounded (Holo)) exhibit significantly improved localization compared to those identified by traditional pocket-hunting tools (GT + Pocket Detection) or predicted apo structures (Predicted Unbounded (Apo)) (Fig.~\ref{fig:fig2}d). Specifically, 48.0\% of HelixFold3 holo predictions fall within the Near-native category, compared to 27.0\% for ground-truth structures with pocket detection and 0.0\% for predicted apo structures. Conversely, predicted apo structures showed 83.0\% of predictions in the Distant category, highlighting the difficulty of defining binding pockets in the absence of ligands. These results indicate that the inaccuracy of traditional pocket identification constitutes a primary bottleneck in decoupled workflows, whereas the end-to-end approach inherently resolves this by simultaneously localizing the ligand and the protein during the joint folding and docking process.

\begin{figure}[htb!]  
    \centering
    \includegraphics[width=0.98\linewidth]{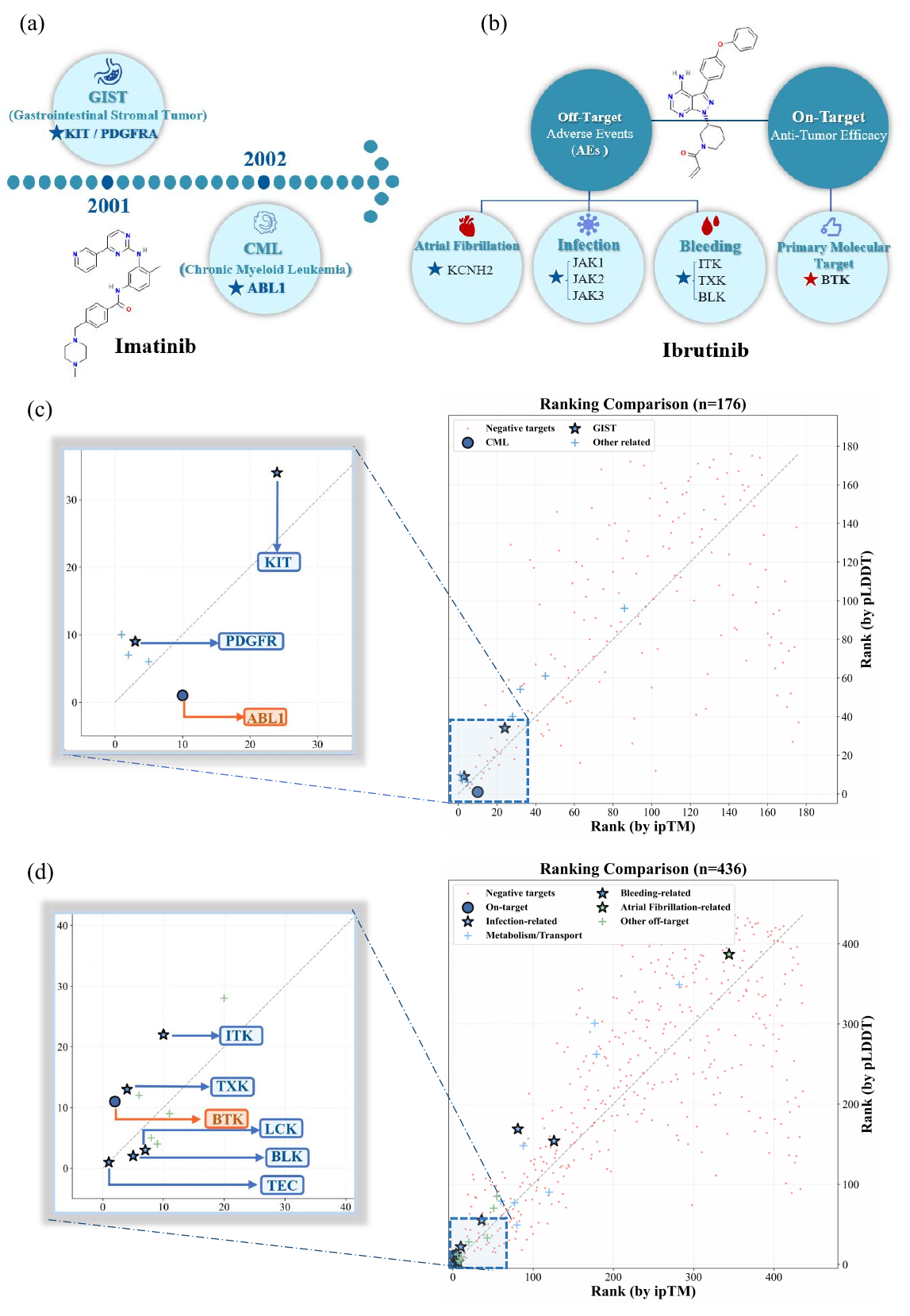}
    \caption{\textbf{Application framework and case studies for reverse screening.} \textbf{(a)} Imatinib case background: timeline and therapeutic targets (GIST: KIT/PDGFRA; CML: ABL1). \textbf{(b)} Ibrutinib case background: on-target anti-tumor efficacy (BTK) and off-target adverse events (bleeding: ITK, TXK, BLK; infection: JAK family; atrial fibrillation: KCNH2). \textbf{(c)} Imatinib case study: ranking comparison between ipTM and pLDDT metrics. \textbf{(d)} Ibrutinib case study: ranking comparison between ipTM and pLDDT metrics.}
    \label{fig:fig4}
\end{figure}

\subsection{Retrospective Validation in Drug Repurposing} \label{Imatinib}


To evaluate the practical utility of the HelixFold3-based end-to-end strategy, we conducted a blind retrospective screening using Imatinib as a classic case study (Fig.~\ref{fig:fig4}a, c). Originally developed as a selective BCR-ABL inhibitor for its first indication, chronic myeloid leukemia (CML), Imatinib was later successfully repurposed for its secondary indication, gastrointestinal stromal tumors (GIST), following the discovery of its potent activity against KIT and PDGFRA kinases \cite{imatinibsynergy2014}.

Following the standardized workflow described in Section~\ref{sec:methods} (Methods), we constructed a target library of 176 proteins through multi-source database integration, which included 10 validated binding targets: one primary driver for the first indication (ABL1), two key targets for the secondary indication (KIT and PDGFRA), and seven additional validated binding partners (RET, CSF1R, DDR1, DDR2, PDGFRB, BCR, and MCL1) that represent structurally related kinases within the kinome. We implemented two parallel ranking strategies based on HelixFold3’s internal confidence metrics: ipTM and pLDDT. Both metrics demonstrated robust discriminative power, successfully prioritizing therapeutic targets against a background of 166 negative targets (non-binding decoys and structural homologs), as summarized in Fig.~\ref{fig:fig4}c. The distribution of known targets by indication category is as follows: one target for the primary indication (CML), two targets for the secondary indication (GIST), and seven additional related targets. Fig.~\ref{fig:fig4}c presents a two-level visualization strategy: the main scatter plot displays the overall clustering effect of all 176 targets, revealing the global distribution pattern of target rankings across the candidate library, while the inset provides a magnified view focusing on the precise positions of the most critical therapeutic targets. In this visualization, filled circles denote the primary on-target (ABL1 for CML), whereas star symbols represent key therapeutic targets for the secondary indication (KIT and PDGFRA for GIST) and other validated binding partners. We note that the designations "on-target" and "off-target" refer to therapeutic context and clinical relevance, not to positive versus negative samples in the experimental validation sense. Specifically, on-targets are the primary intended therapeutic targets, while off-targets (or secondary targets) are additional validated binding partners that may contribute to efficacy or adverse effects; both represent experimentally confirmed positive binding interactions. Notably, both ipTM and pLDDT identified 5 of these 10 validated binding partners within the Top-10 predictions, achieving a 50.0\% Top-10 success rate. Regarding overall coverage, the ipTM-based strategy successfully recovered all primary therapeutic targets (ABL1, KIT, and PDGFRA) within the Top-30, while the pLDDT-based strategy captured the same set within the Top-40. Specifically, when using pLDDT as the scoring metric, ABL1 was ranked first (rank 1/176), demonstrating perfect identification of the first-indication target. When using ipTM as the scoring metric, PDGFRA achieved the best ranking (rank 3/176), followed by ABL1 (rank 10/176) and KIT (rank 24/176). The interface confidence metrics for these predictions were consistently high (ipTM $>$ 0.85, pLDDT $>$ 0.70), with ABL1 showing particularly high confidence (ipTM $=$ 0.9584, pLDDT $=$ 93.81). These results, which include the successful prioritization of auxiliary kinases such as CSF1R (rank 1/176 by ipTM) and DDR1/2 (ranks 5/176 and 2/176 by ipTM), underscore the transformative potential of end-to-end reverse screening in reconstructing the comprehensive polypharmacological landscape of existing drugs.

\subsection{Retrospective Investigation of Off-Target Prediction} \label{Ibrutinib}

To further evaluate the predictive utility of the HelixFold3-based end-to-end strategy in deciphering complex safety profiles, we applied the framework to Ibrutinib, a covalent Bruton's tyrosine kinase (BTK) inhibitor (Fig.~\ref{fig:fig4}b, d). While highly effective for B-cell malignancies, Ibrutinib exhibits significant polypharmacology, where off-target interactions drive adverse clinical events such as bleeding, infections, and atrial fibrillation \cite{ibrutinibofftarget2021}. Following the same standardized workflow (see Section~\ref{sec:methods}), we constructed a candidate library of 436 proteins, which integrated 26 validated targets stratified into categories: one primary on-target (BTK), four bleeding-related off-targets (TEC, TXK, BLK, and ITK), five infection-related off-targets (LCK, ITK, JAK1, JAK2, and JAK3), and 16 auxiliary binding or metabolic partners.

Using the same dual ranking strategy (ipTM and pLDDT), we prioritized key therapeutic and off-targets against a background of 410 negative targets, as summarized in Fig.~\ref{fig:fig4}d. The distribution of known targets by adverse event category is as follows: one on-target, four bleeding-related off-targets, five infection-related off-targets, one Atrial Fibrillation-related off-target, and additional metabolism/transport and other off-targets. Fig.~\ref{fig:fig4}d employs the same two-level visualization strategy: the main scatter plot shows the overall clustering effect of all 436 targets, and the inset magnifies the critical therapeutic and adverse-event-related targets. In this visualization, filled circles denote the primary on-target (BTK), whereas star symbols represent off-targets associated with specific adverse events, with different colors indicating different categories (bleeding-related, infection-related, and Atrial Fibrillation-related). HelixFold3 identified the on-target BTK with high precision (ipTM rank 2/436, pLDDT rank 11/436). Remarkably, the ipTM-based strategy achieved a 100\% Top-10 success rate for bleeding-related off-targets, precisely capturing TEC (rank 1/436), TXK (rank 4/436), BLK (rank 5/436), and ITK (rank 10/436). This exceptional performance reflects the model’s ability to recognize the structural conservation within the Tec family kinases, which share the active-site architecture required for Ibrutinib binding. For infection-related targets, the model successfully prioritized LCK (rank 7/436) and ITK within the Top-10, though the more structurally divergent JAK family kinases were ranked lower (JAK2: rank 36/436, JAK3: rank 81/436, JAK1: rank 126/436).

In contrast, the model faced challenges in predicting the Atrial Fibrillation-related ion channel KCNH2 (hERG), which ranked 344th (ipTM) and 387th (pLDDT), with low confidence scores (ipTM $=$ 0.3437, pLDDT $=$ 48.80). This disparity highlights a current limitation of end-to-end models in handling complex transmembrane proteins compared to soluble kinases. Overall, the ipTM-based strategy identified 34.6\% (9/26) of all known targets within the Top-10, demonstrating a particularly strong enrichment for the Src and Tec kinase families. These results underscore the transformative potential of end-to-end reverse screening in reconstructing the comprehensive polypharmacological landscape of existing drugs, providing a powerful approach for predicting off-target liabilities and assessing safety profiles during drug development.

\section{Discussion}

Our end-to-end reverse screening framework, built upon HelixFold3, demonstrates significant advantages over traditional step-wise reverse docking approaches. The unified modeling of protein folding, ligand docking, and affinity estimation eliminates error propagation inherent in step-wise workflows, resulting in substantially improved screening accuracy and structural fidelity. Recent advances in end-to-end protein-ligand complex prediction, including NeuralPLexer \cite{neuralplexer2024}, RoseTTAFold All-Atom \cite{rosettafold2024}, and other deep learning-based methods \cite{ligpose2024,diffdock2022,e3bind2022}, have shown promise in directly predicting complex structures from sequences. However, these methods have primarily focused on docking accuracy rather than large-scale reverse screening applications. The benchmark evaluation reveals that our approach achieves a Top-1\% success rate of 38.0\%, representing a several-fold improvement over conventional methods even when those baselines are granted access to crystal protein structures and pocket coordinates. This performance leap is underpinned by superior accuracy across all dimensions: target structure prediction, binding site localization, ligand docking, and affinity scoring.


The case studies on Imatinib and Ibrutinib demonstrate the framework's practical utility in addressing real-world drug discovery challenges. These retrospective validations showcase our ability to construct comprehensive candidate target libraries through standardized workflows that integrate multiple authoritative databases and systematically expand target lists based on application scenarios. The Imatinib case study, representing drug repurposing, successfully identified all primary therapeutic targets within the Top-40, achieving a 50.0\% Top-10 success rate. The Ibrutinib case study, representing off-target prediction, demonstrated exceptional performance for structurally related kinase families, achieving a 100\% Top-10 success rate for bleeding-related off-targets. These case studies reflect our framework's generalizability and capability to construct realistic problem scenarios that capture the complexity of real-world drug discovery applications, from identifying new therapeutic indications to predicting adverse effects.


While our framework demonstrates substantial improvements, several limitations remain. The challenges in predicting ion channel targets, as observed with KCNH2 (hERG), indicate that complex transmembrane proteins require further investigation. Additionally, the method's generalization to highly novel structures may benefit from expanded training data or architectural improvements. Future work should focus on addressing these limitations while maintaining the framework's scalability and ease of use.

In conclusion, our end-to-end reverse screening strategy provides a robust and scalable platform for investigating molecular mechanisms, exploring off-target interactions, and supporting rational drug discovery. The standardized workflow for constructing candidate target libraries, combined with the superior performance demonstrated across diverse evaluation scenarios, establishes this framework as a valuable tool for both fundamental research and practical drug discovery applications.

\section{Methods}\label{sec:methods}

\subsection{Evaluation Datasets}\label{sec:datasets}

\begin{figure}[t]
\centering
\includegraphics[width=1.0\linewidth]{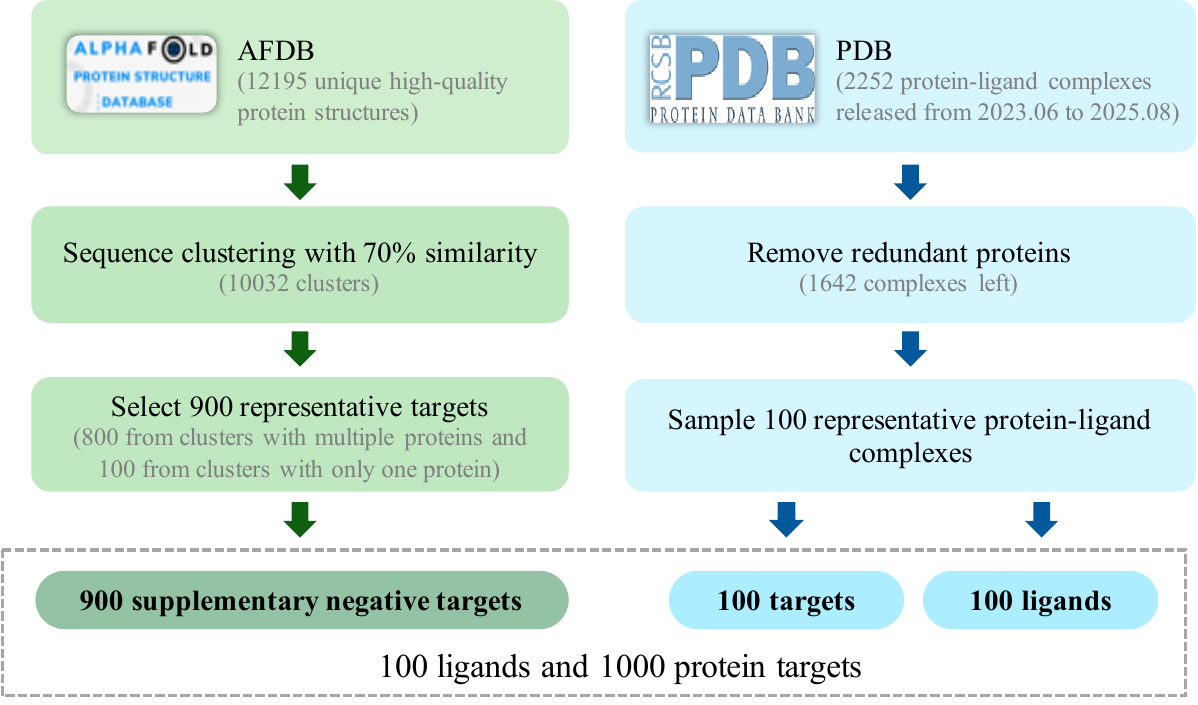}
    \caption{\textbf{Benchmark dataset construction pipeline.} Representative negative protein targets selected from AFDB  (left) and positive protein-ligand pairs selected from recent PDB protein-ligand complexes (right).}
\label{fig:datapipline}
\end{figure}

To establish a comprehensive evaluation framework for reverse screening, we constructed a diverse benchmark dataset comprising both positive and negative samples (Fig.~\ref{fig:datapipline}). The detailed process is introduced in the following sections.

\paragraph{Negative Sample Selection}
For negative sample generation (left pipeline in Fig.~\ref{fig:datapipline}), we selected samples from the AFDB \cite{alphafolddb2022} human proteome to provide a comprehensive background for reverse screening evaluation. Following the methodology established by Luo et al. \cite{luo2024benchmarking}, we began with 23,586 human protein structures from the AFDB.

\textbf{Quality Control for Structures from AFDB.} To ensure the quality and suitability of structures for reverse screening evaluation, we implemented a rigorous quality control pipeline. The quality control process primarily focused on structural quality assessment and sequence length constraints. Low-quality structures, including disordered protein structures and those lacking well-defined ligand binding pockets, were systematically removed. Additionally, we applied a sequence length filter to exclude proteins exceeding 1,500 amino acids, ensuring computational tractability while maintaining structural diversity. This initial filtration process reduced the dataset from 23,586 structures to 14,424 high-quality AlphaFold2 \cite{jumper2021highly} (AF2) predicted structures.

To further validate structural reliability, we compared these filtered AF2 predicted structures against experimental structures from the PDB \cite{berman2000protein} using DeepAlign \cite{wang2013deepalign} to compute structural similarity (TM-score \cite{zhang2004tmscore}). For proteins with multiple experimental structures, we selected the structure with the lowest resolution for comparison. Only structures exhibiting TM-score greater than 0.8 were retained, resulting in 12,195 unique high-quality protein targets suitable for clustering analysis.

\textbf{Clustering and Representative Selection.} Starting from these 12,195 unique proteins, we performed sequence-based clustering analysis using MMseqs2 \cite{steinegger2017mmseqs2} with a 70\% identity threshold, resulting in representative clusters that capture the sequence diversity of the human proteome. The clustering process yielded 10,032 single-protein clusters and 800 multi-protein clusters. From these clusters, we selected 900 representative proteins to serve as negative samples using a two-stage selection strategy. First, we selected one representative protein from each of the 800 multi-member clusters, ensuring comprehensive coverage of protein families and functional domains. Second, we randomly sampled 100 proteins from the 10,032 singleton clusters to include unique protein structures that lack close homologs. This selection strategy balances diversity with computational feasibility, providing a robust negative control set that represents the full spectrum of human protein structures while maintaining manageable computational requirements for large-scale evaluation.

\paragraph{Positive Sample Collection}
For positive sample generation (right pipeline in Fig.~\ref{fig:datapipline}), we derived samples from recently published protein-ligand complexes in the Protein Data Bank (PDB) \cite{berman2000protein}, ensuring structural novelty and relevance to current research. We collected 5,780 PDB structures deposited between June 2023 and August 2025, representing 456 unique species. Following species-based filtering, we identified 2,252 human protein structures corresponding to 1,642 unique protein structures.

\textbf{Ligand Quality Assessment and Selection.} The ligand selection process prioritized high-quality small molecules with well-characterized binding interactions. We extracted three-dimensional ligand structures directly from experimental protein-ligand complexes, ensuring that the binding conformations were experimentally validated. This approach eliminates the uncertainty associated with ligand conformation generation and provides a more reliable foundation for reverse screening evaluation.

For each protein-ligand pair, we implemented a multi-dimensional scoring system that considered protein sequence identity, ligand diversity, ligand quality metrics (molecular weight 150-800 Da, atom count 10-80, LogP 0-5, rotatable bonds $<$ 10), and protein length (100-500 amino acids optimal). This comprehensive scoring approach ensured the selection of high-quality, diverse, and computationally tractable protein-ligand pairs.

\textbf{Quality Control and Validation.} We implemented a multi-stage quality control pipeline to ensure the reliability and scientific validity of the final benchmark dataset. The quality control process consisted of two sequential validation stages applied to the sampled protein-ligand pairs.

In the first stage, we performed direct Vina scoring and re-docking validation on the experimental PDB \cite{berman2000protein} conformations using AutoDock Vina \cite{trott2010autodock,vina2021}. Each protein-ligand pair underwent two validation modes: score-only evaluation (assessing the binding affinity of the experimental conformation) and re-docking validation (testing the ability to reproduce the experimental binding pose). We employed ligand-centric pocket detection with a 5.0 Å wrapping distance around the experimental ligand coordinates, ensuring appropriate binding site definition. Pairs that failed processing or exhibited abnormal score-only values ($> -0.1$  kcal/mol, indicating potential structural issues) were systematically removed.

In the second stage, we subjected the remaining protein-ligand pairs to HelixFold3 conformation generation and prediction. The predicted structures were then evaluated for structural quality, and pairs exhibiting problematic characteristics were excluded. Specifically, we removed conformations with highly structurally redundant conformations, multimeric assemblies (polymer assemblies), missing key ions or cofactors, and those with extended unresolved regions. This comprehensive quality control process resulted in a final dataset of 100 high-quality protein-ligand pairs, each representing a unique target-ligand interaction with validated binding characteristics and predicted structural consistency.

\paragraph{Final Benchmark Dataset}

The final benchmark dataset comprises a comprehensive reverse screening evaluation set consisting of 100 high-quality ligands paired against 1,000 proteins (100 positive sample proteins + 900 negative sample proteins), resulting in 100,000 ligand-protein pairs for evaluation. The positive sample ligands were selected through a multi-dimensional scoring system that considers protein sequence identity, ligand diversity, and structural quality.

This comprehensive dataset design enables evaluation of reverse screening performance across diverse structural and chemical space. Each of the 100 ligands is tested against all 1,000 proteins, allowing for assessment of both true positive identification (ligand binding to its known target) and false positive rejection (ligand binding to non-target proteins).

\subsection{Reverse Screening Pipelines}
For reverse screening, we compared our HelixFold3-based end-to-end pipeline with a baseline reverse docking workflow using AutoDock Vina \cite{trott2010autodock,vina2021}. In the HelixFold3 pipeline, proteins from the protein library were provided as amino acid sequences, and small-molecule ligands were represented in SMILES format. The HelixFold3 API\footnote{https://paddlehelix.baidu.com/app/tut/guide/all/helixfold3sdk} automatically performs Multiple Sequence Alignment (MSA) retrieval, structure prediction, and protein–ligand complex modeling, producing the predicted structures along with confidence metrics (ipTM and pLDDT). To further refine the ranking of predicted interactions, we applied AutoDock Vina for affinity evaluation using the HelixFold3-predicted complex structures. Specifically, we used the ligand coordinates from the predicted complexes to define binding pockets, and then performed Vina scoring to assess the binding affinity of these predicted structures. The reverse screening computational pipeline processes targets and ligands from the combined dataset (100 ligands, 1,000 targets) through HelixFold3 for structure prediction and confidence scoring (ipTM and pLDDT), with additional Vina rescoring applied to predicted structures, ultimately producing ranked target lists based on binding affinity and confidence metrics.

In contrast, the baseline Vina pipeline required three-dimensional protein structures, which were obtained either from experimental PDB \cite{berman2000protein} entries or predicted by AF2 \cite{jumper2021highly}. Small-molecule ligands were provided as three-dimensional SDF structures extracted from experimental protein-ligand complexes in the PDB \cite{berman2000protein}, which were processed using RDKit \cite{bento2020open} and Meeko \cite{meeko2024}, and subsequently prepared in PDBQT format for docking. Candidate pockets were identified using fpocket \cite{le2009fpocket}  and P2Rank \cite{krivak2018p2rank}, followed by Vina docking and scoring based on the predicted binding affinity. This workflow highlights the differences in input requirements and methodology between sequence-based end-to-end modeling and traditional structure-driven reverse docking approaches.

\subsection{Target Library Construction for Real-World Applications}

For real-world reverse screening applications (described in Section \ref{Imatinib} and Section \ref{Ibrutinib}), we established a standardized workflow for constructing candidate target libraries that integrates multiple authoritative databases including FDA labels, DrugBank \cite{knox2024drugbank}, BindingDB \cite{gilson2015bindingdb}, PubChem \cite{kim2024pubchem}, ChEMBL \cite{zdrazil2024chembl,davies2015chemblwebservices}, and KEGG \cite{kanehisa2000kegg}. The workflow begins with comprehensive data collection from these sources to identify known targets, approved indications, and documented adverse effects. Based on the specific application scenario—whether drug repurposing or off-target prediction—we then systematically expand the candidate target list using ChEMBL. For drug repurposing scenarios, we search for human protein targets associated with relevant disease names or clinical indications. For off-target prediction scenarios, we employ symptom-related keywords (e.g., bleeding, platelet aggregation, infection, immunosuppression, atrial fibrillation, arrhythmia) to identify potential off-target interactions. The ChEMBL-derived targets are then combined with targets identified from the initial database searches, with priority given to experimentally validated targets from BindingDB and DrugBank to serve as ground truth for evaluation. This integrated approach enables the construction of comprehensive candidate target panels that balance known therapeutic targets with expanded candidate sets, providing a robust foundation for reverse screening evaluation across diverse application scenarios.

\subsection{Evaluation Metrics}

To comprehensively evaluate the performance of reverse screening methods, we employed multiple metrics that assess both the accuracy of target identification and the ranking quality of predicted interactions.

The Top-K success rate (SR@K) measures the proportion of ligands for which the true target is correctly identified within the top K ranked predictions. For each ligand, we ranked all candidate proteins based on the scoring function (e.g., ipTM, pLDDT, or Vina score), and determined whether the true target appeared within the top K positions. The success rate is calculated as:

\begin{equation}
\text{SR@K} = \frac{N_{\text{top-K}}}{N_{\text{total}}}
\end{equation}

where $N_{\text{top-K}}$ is the number of ligands whose true targets are ranked within the top K, and $N_{\text{total}}$ is the total number of ligands in the benchmark. We evaluated performance at multiple K values: Top-1, Top-10, Top-20, Top-50, and Top-100. Additionally, we computed Top-1\% success rate, which corresponds to the top 1\% of ranked targets (i.e., top 10 targets out of 1,000 candidates), providing a more stringent evaluation criterion that accounts for the large search space.

\bibliography{sn-bibliography}

\end{document}